\documentclass[pra,twocolumn]{revtex4-2}
\usepackage{xcolor}
\usepackage{todonotes}
\usepackage{amsmath,amssymb,bm, bbm}
\usepackage{physics}
\usepackage{upgreek}
\usepackage{graphicx}
\usepackage{dcolumn}
\usepackage{dsfont}
\usepackage{nccmath}
\usepackage[colorlinks=true, allcolors=blue]{hyperref}
\usepackage{amsthm}
\usepackage{units}
\usepackage{mathrsfs}
\usepackage{hhline}
\usepackage{subfigure}
\usepackage{svg}
\usepackage{soul}
\usepackage{enumitem}
\setlist[enumerate]{itemsep=0mm}

\begin{document}
\title{Quantum-Enhanced Parameter Estimation Without Entanglement}
\author{Pragati Gupta}
\email{pragati.gupta@ucalgary.ca}
\affiliation{Institute for Quantum Science and Technology, University of Calgary, Alberta T2N 1N4, Canada}

\begin{abstract}
Entanglement is generally considered necessary for achieving the Heisenberg limit in quantum metrology. 
We construct analogues of Dicke and GHZ states on a single $N+1$ dimensional qudit 
that achieve precision equivalent to symmetrically entangled states on $N$ qubits, showing that entanglement is not necessary for going beyond the standard quantum limit. 
We define a measure of non-classicality based on quantum Fisher information and  estimate the achievable precision, suggesting a close relationship between non-classical states and  metrological power of qudits.
Our work offers an exponential reduction in the physical resources required for quantum-enhanced parameter estimation, making it accessible on any quantum system with a high-dimensional Hilbert space.    
\end{abstract}

\maketitle

The precise measurement of physical quantities, from electromagnetic fields, to temperature and pressure, is important, both for applications like detection of gravitational waves, and for fundamental aspects like phase sensitivity in interferometry~\cite{DRC17RevModPhys}.
The standard quantum limit (SQL), which arises from the discrete nature of quantum measurements, describes the limit of precision achievable in parameter estimation using a system of $N$ independent quantum probes.
In contrast to uncorrelated probes, correlated many-body systems such as squeezed states can improve measurement precision beyond the SQL by increasing the precision of a chosen quantum observable at the expense of the uncertainty of another conjugate observable~\cite{GLM04Science,GLM06PRL,GLM11Natphotonics}.
Quantum sensing assisted by entangled probes, like the Greenberger-Horne-Zeilinger (GHZ) states, $N00N$ states or Dicke states  can achieve the Heisenberg limit~\cite{PSO18RMP,CPA+22NatPhys,TA14JoPA,DPG23PRB}.

Quantum Fisher information (QFI), the quantum analogue of the classical Fisher information, plays a central role in quantum metrology, where its inverse determines the achievable precision, given by the quantum Cram\'er-Rao bound~\cite{TA14JoPA}.  
The value of QFI  for a quantum system scales with the effective number of sub-systems, motivating its use as measure of macroscopicity, as well as a quantifier for the degree of separation between two components of a macroscopic Schr\"odinger cat state, using the so-called the relative QFI  \cite{FSD+18RevModPhys, FD12NJP}.
Recently, QFI was established as a measure of non-classicality of a state in continuous variable systems, suggesting that metrological power of a system is related to the degree of macroscopicity~\cite{KTVJ19PRL}. 
Operational resource-theoretic definitions of metrological power, non-classicality and macroscopicity can also be obtained based on QFI, making it a powerful tool for experimental investigations~\cite{GJA20PRR}.
Further, QFI universally captures resourcefulness of any quantum state, regardless of the specific unitary parameter encoding, hence, is important for general quantum resource theories~\cite{TNR21PRL}.

While entangled states can maximize QFI, they are difficult to create and maintain, due to the high-degree of required control and their susceptibility to decoherence~\cite{HMP+97PRL}.
High-dimensional systems offer an alternate hardware-efficient way of using a multi-level structure, naturally available on most quantum systems, for storing and processing of information and reducing circuit complexity~\cite{WHSK20FrontPhys}.
The high-dimensional Hilbert space available within a harmonic oscillator can also be utilized to encode quantum information on non-classical states and perform error-correction~\cite{PJG+20ScienceAdv}. 
Motivated by the importance of qudits in above examples, and the fact that many quantum sensing technologies such as  superconducting systems~\cite{SZS+18PRA} and NV centres~\cite{DFD+11NatPhys} host natural high-dimensional Hilbert spaces, we ask if qudits can serve as a resource for quantum metrology.

We introduce ``qudit-assisted" protocols for parameter estimation that attain an enhanced precision---scaling linearly with the dimension---without using entanglement.
The quantum Fisher information (QFI) for a $d$-dimensional qudit is equivalent to that of a $N=d-1$ multi-qubit system under exchange symmetry~\cite{ZSM+13PRA}. 
We define resource theoretic measures of metrological gain and non-classicality for a qudit, relating them to the gain in precision and analyze the effect of decoherence as the system size increases. 
We construct analogues of Dicke states and GHZ state on a qudit---respectively the orthonormal basis states and the superposition of the lowest and highest eigenvalue state---and analytically show that metrological protocols with these states attain the Heisenberg limit. 
Our work suggests that an exponential reduction in the state space dimension---from $2^{N}$ to $N+1$---is achievable in metrology using qudits, which could make quantum-enhanced parameter estimation accessible with independent probes.

This paper is organized as follows. In Sec.~\ref{sec:QFI}, we discuss quantum Fisher information of a qudit, its equivalence to a multi-qubit state with permutation symmetry, and define measures for non-classicality and metrological power. In Sec.~\ref{sec:protocols}, we describe metrological protocols and analyze the achieved precision with Dicke-like states (\ref{sec:dicke}) and GHZ-like states (\ref{sec:GHZ}) on a qudit. We analyze the effects of system size on decoherence rates in Sec.~\ref{sec:decoherence} and finally, give concluding remarks in Sec.~\ref{sec:conclusion}.

\section{Theoretical Measures for a Qudit}\label{sec:QFI}

\subsection{Quantum Fisher information}
Quantum Fisher information (QFI) is the quantum analogue of classical Fisher information
and used to quantify the rate of change of a state's density matrix $\rho$
with respect to an unknown parameter $\theta$.
Quantum Fisher information, $F_{\text Q}$,  is defined in terms of the symmetric logarithmic derivative (SLD) $\hat L_\theta$,
which is a Hermitian operator given by
\begin{equation}\label{eq:SLD}
    \partial_\theta \rho := \frac{1}{2}\{\rho,\hat L_\theta\},
\end{equation}
where $\partial_\theta\equiv \nicefrac{\partial}{\partial \theta}$ and $\{.,.\}$ denotes the anticommutator. 
In terms of the SLD operator, $F_{\text Q}$ is given by
\begin{equation}\label{eq:QFI-SLD}
    F_{\text Q} := \text{Tr}\left(\rho \hat L_\theta^2 \right) 
    = \text{Tr}\left [(\partial_\theta \rho)\hat L_\theta \right],
\end{equation}
which defines QFI without explicit diagonalization of the density matrix.

We calculate the QFI of  a $d$-level system using the generalized Bloch sphere representation~\cite{ZSM+13PRA}. 
We use the generalized Gell-Mann matrices $\bm E = \{E_i\}_{i=1}^{d^2-1}$ that generate the Lie algebra corresponding to the special unitary  group SU$(d)$~\cite{BK08JoPAMT} and the Bloch vector $\bm\omega\in\mathbb{R}^{d^2-1}$ for denoting the density matrix as 
\begin{equation}\label{eq:bloch}
    \rho = \frac{1}{d}\mathbb{I}_d + \frac{1}{2}\bm \omega\cdot\bm E,
\end{equation}
where $\mathbb{I}_d$ is a $d$-dimensional identity matrix. 
For pure states, $\rho = \rho^2$, and we can write
\begin{equation}
    \partial_\theta \rho = \partial_\theta \rho^2 = \rho\left(\partial_\theta \rho\right) + \left(\partial_\theta \rho\right)\rho.
\end{equation}
Comparing the above equation with (\ref{eq:SLD}), we get
\begin{equation}
    \hat L_\theta = 2\partial_\theta \rho,
\end{equation}
for pure states. This can be substituted in (\ref{eq:QFI-SLD}) to get
\begin{equation}\label{eq:QFI-pure-bloch}
    F_\text Q = \text{Tr}\left[(\partial_\theta \rho)2\partial_\theta \rho\right] = |\partial_\theta \bm \omega|^2,
\end{equation}
where the second equality is obtained from (\ref{eq:bloch}) and the relation $\text{Tr}\left[(\bm a\cdot\bm E)(\bm b\cdot\bm E)\right] = 2\bm a\cdot\bm b$. 
Equation \ref{eq:QFI-pure-bloch} holds for a two-level system, i.e.~$d=2$, where the density matrix is expressed in terms of Pauli matrices $\bm \sigma = \{\hat\sigma_x,\hat\sigma_y,\hat\sigma_z\}$, which generate the Lie algebra for SU(2) group.

QFI (Eq.~\ref{eq:QFI-SLD}) can also be expressed in terms of the eigenvalues $\lambda_i$ and eigenvectors $\psi_i$ of the density matrix the observable $\hat A$ as
\begin{equation}\label{eq:}
    F_{\text Q}[\rho,\hat A] = 
    2\sum_{i\neq j} \frac{\left(\lambda_i -\lambda_j\right)^2}{\lambda_i + \lambda_j}|\langle \psi_i|\hat A|\psi_j\rangle |^2,
\end{equation}
where $\lambda_i + \lambda_j>0$. For pure states, the density matrix $\rho = \ket{\psi}\bra{\psi}$ and the eigenvalues are all zero except for the state $\ket{\psi}$.  The above equation can be written as 
\begin{equation}
\begin{split}
    F_\text Q = 
    &2\sum_{\lambda_i=0,\lambda_j\neq0} \frac{(0-1)^2}{0+1} |\langle \psi_i | \hat A | \psi\rangle |^2 \\
    &+ 2 \sum_{\lambda_i\neq0,\lambda_j=0} \frac{(1-0)^2}{1+0} |\langle \psi | \hat A | \psi_i\rangle |^2\\
    = &4 \sum_{\lambda_i=0}  |\langle \psi_i | \hat A | \psi\rangle |^2
    = 4 \sum_{\lambda_i=0}  \langle \psi | \hat A | \psi_i\rangle\langle \psi_i | \hat A | \psi\rangle.
\end{split}
\end{equation}
Using $\sum_{\lambda_i =0}| \psi_i\rangle\langle \psi_i | = \mathbb{I}-|\psi\rangle\langle \psi|$,
the QFI reduces to 
\begin{equation}
    F_\text Q (\psi,\hat A) = 4\left[ \langle \psi|\hat A^2|\psi \rangle - |\langle\psi|\hat A|\psi\rangle|^2 \right].
\end{equation}
Thus, for pure states, the QFI is four times the variance of the state $\psi$ with respect to observable $\hat A$. Using convexity of QFI~\cite{GJA20PRR},
\begin{equation}\label{eq:convexity}
    F_\text Q [\rho,\hat A] \leq \sum_i \lambda_i F_\text Q [\psi_i,\hat A], 
\end{equation}
where $\lambda_i $ 
and $\psi_i$ 
are eigenvalues and eigenstates of the density matrix, as above. 
For mixed states, QFI is
\begin{equation}\label{eq:QFI-mixed}
    F_\text Q [\rho,\hat A] = \min _{\{\lambda_i,\psi_i\} } \sum_i \lambda_i F_\text Q (\psi_i,\hat A),
\end{equation}
which is four times the convex roof of variance, minimized over eigen decompositions of the density matrix.

\paragraph*{Equivalence to symmetrically entangled states:} Quantum Fisher information of a multi-qubit system with permutational symmetry is equivalent to that of a qudit. To see this, we express a multi-qubit operations using the collective spin operators $\bm J = \{\hat J_x, \hat J_y, \hat J_z\}$,
\begin{equation}
    \hat J_i = \sum_{n=1}^N \frac{\sigma_i^n}{2}; \{ \hat J_i, \hat J_j \} = i\epsilon_{ijk} \hat J_k, 
\end{equation}
where $N$ is the total number of qubits, $i\in\{x,y,z\}$ and $\sigma_i^n$ acts on the $n^{\text{th}} $ qubit and
$\hat J_i$ are the $N+1$ dimensional representations of the SU(2) group. 
Since, SU(2) is a subgroup of the SU($N+1$) group, collective operations on permutationally symmetric qubits can be mapped to operations on a $d$-dimensional qudit. 
Similarly, the $N+1$ levels of a qudit correspond to multi-qubit Dicke states in the $\hat J_z$ basis~\cite{D54PR}, which are symmetrically entangled states, and this mapping can be used to construct analogous metrological probes on a qudit.

\subsection{Non-classicality}
Quantum Fisher information is closely related to the macroscopicity of a spin state~\cite{FD12NJP, FSD+18RevModPhys}. For multi-qubit systems, QFI can be used to quantify the effective number of systems forming a quantum state
\begin{equation}
    N_\text{eff}(\rho) = \max_{\hat A} \frac{F_\text Q(\rho,\hat A)}{N},
\end{equation}
which has a range $1\leq N_\text{eff}(\rho)\leq N$ for pure states. A multi-qubit state is called macroscopic if $N_\text{eff}$ is linear in the system size. Macroscopicity of a qudit can similarly be defined as the effective number of degrees of freedom in a high-dimensional state
\begin{equation}
    d_\text{eff}(\rho) = \max_{\hat A} \frac{F_\text Q(\rho,\hat A)}{d-1},
\end{equation}
where $d$ is the dimension of the qudit, and $\hat A$ is any linear operator. We note that for pure states, $1\leq d_\text{eff}(\rho)\leq d-1$ under a linear parameter encoding.

The notion of macroscopicity is closely related to the non-classicality of a state. To define non-classicality $\mathcal{N}(\rho)$ from a resource theory perspective, we note that any such measure should satisfy four conditions~\cite{GJA20PRR,KTVJ19PRL}: non-negativity, weak monotonicity, strong monotonicity and convexity. Here, non-negativity means that $\mathcal{N}(\rho)\geq 0$ and the equality is satisfied if and only if $\rho$ is a classical state. Weak monotonicity means that $\mathcal{N}$ cannot be increased by the application of a classical operation. In the context of a system comprising of qubits, classical operations correspond to linear rotations of one or more qubits around the Bloch sphere. For qudits, classical operations could be defined as linear operations that transform a SU(d) coherent state into another SU(d) coherent state~\cite{N00JoPAMath}. Strong monotonicity means that $\mathcal{N}$ should not increase when a subset of the system is measured, but, this condition is not relevant for the discussion considering only a single qudit, and convexity denotes. Based on the above requirements, we define non-classicality of a qudit to be
\begin{equation}
    \mathcal{N}(\rho) = 
    \min_{\{\lambda_i,\psi_i\} }
    \left[ \max_{\hat A} 
        \left(\sum_i \lambda_i \text{Var}(\psi,\hat A)\right)
    \right] - (d-1),
\end{equation}
where, Var$(\psi,\hat A) =\left[ \langle \psi_i|\hat A^2|\psi_i \rangle - |\langle\psi_i|\hat A|\psi_i\rangle|^2 \right]$ and the minimization is over different decompositions of the density matrix . An entanglement measure for permutationally symmetric multi-qubits can be obtained by substituting $d=N+1$, where $N$ is the number of qubits.

\subsection{Metrological power}
The quantum Cram\'er-Rao bound is the quantum analogue of the classical Cram\'er-Rao bound and determines the precision $\Delta \theta$ achievable with a quantum probe,
\begin{equation}\label{eq:cramer-rao}
    (\Delta \theta)^2 \geq \frac{1}{m F_\text Q [\rho,\hat A]},
\end{equation}
where $m$ is the number of measurements. For a pure two-level system with $F_\text Q =4$, or, $N$ uncorrelated two-level systems with $F_\text Q =4N$, the above equation gives the well-known standard quantum limit. 
Entangled systems go beyond this limit and  have $F_\text Q \leq 4 N^2$, with GHZ states saturating the bound.
From a resource theory perspective, a measure of metrological power, $\mathcal{W}(\rho)$, should be non-negative and $\mathcal{W}(\rho)>0$ should imply precision beyond that standard quantum limit.
Similar to the approach used for continuous-variable systems~\cite{GJA20PRR}, we define  metrological power $\mathcal{W}(\rho)$ as the amount by which the QFI of a state is greater than the maximum QFI for any classical state 
\begin{equation}\label{eq:relationWN}
    \mathcal{W}(\rho) = \max \left[d_\text{eff}-1,0\right].
\end{equation}
Metrological power is related to the achievable precision with a state, in that it captures its ability to go beyond the standard quantum limit. It is bounded by non-classicality
\begin{equation}
    \mathcal{W}(\rho)\leq \mathcal{N}(\rho),
\end{equation}
where the equality holds only for pure states. It is worth noting that not all non-classical states have metrological power, as mixed states could have $d_{\text{eff}}<1$.

\section{Qudit-Assisted Metrological Protocols}\label{sec:protocols}

In this section, we show explicit examples of quantum-enhanced metrology using non-classical states on a qudit. A metrological protocol consists of the following general scheme~\cite{PSO18RMP}: (i) probe preparation: a quantum system is initialized to a desired state, (ii) parameter encoding: the state is manipulated using a an operator that depends on an unknown parameter $\theta$, (iii) readout: the probe is measured in a way that the expectation value of the observable carries information about the encoded parameter and (iv) estimation: calculating the parameter from measurements. The uncertainty $\Delta \theta$ for a protocol crucially depends on all the four operations and is given by the error propagation formula~\cite{B22PRXQ}
\begin{equation}\label{eq:precision}
    (\Delta\theta)^2 := \frac{(\Delta \hat A)^2}{|\partial_\theta\langle \hat A\rangle|^2},
\end{equation}
where $\langle \hat A \rangle$ is the expectation value of operator $\hat A$ and $(\Delta \hat A)^2$ is the variance. 

\subsection{Dicke-like state}\label{sec:dicke}
Many-body entangled states are generally expressed in the Dicke basis, which offers a collective way of treating the $2^N$ degrees of freedom of $N$ qubits using only a $N+1$ dimensional Hilbert space. Dicke states are denoted in the spin representation as $\ket{J,m_J}$, where $J=\nicefrac{N}{2}$ is the collective spin and $m_J\in[-J,J]$, such that $\hat J_z\ket{J,m_J}=m_J\ket{J,m_J}$. In resource theoretic terms, Dicke states represent $k$-entangled states, where $k=m_J$ for small $m_J$, denoting symmetric superpositions of $k$ particles in the excited state. They also called spin number states as they are the spin analogues to the number states. For qudits, the basis states
\begin{equation}
    \ket{i}\equiv\ket{J_d,-J_d+i}; J_d = \nicefrac{(d-1)}{2},
\end{equation}
where $i\in[0,d-1]$. 
From a metrological perspective, Dicke states with $m_J=0$ for integer spins or $m_J=\pm\nicefrac{1}{2}$ for half-integer spins can lead to achieving the Heisenberg limit~\cite{SM95PRL}. Here, we consider the analogous state $\ket{\nicefrac{(d-1)}{2}}$ for odd dimensional qudit and $\ket{\nicefrac{d}{2}}$ and $\ket{\nicefrac{d}{2} -1}$ for even dimensional qudits. 

\paragraph*{State preparation:} Dicke states are generally hard to achieve on entangled systems, but, are much simpler for qudits: any basis state $\ket{i}$ with $i\in[1,d-2]$ is a non-classical Dicke-like state. State preparation can be achieved either by initializing a qudit directly to one of these states or starting from a classical state. 
When the initial state is a coherent state, $\ket{0}$ or $\ket{d-1}$, a non-classical Dicke-like state can be prepared by applying Givens rotations, that selectively address a particular transition, between the initial state and the target state.
This can be achieved with multi-photon transitions or by applying a series of single-photon transitions that lead to the final state.
However, we note that a linear operation alone is not sufficient to initialize to a non-classical Dicke-like state, as it would simply rotate the coherent state to another coherent state.
For Givens rotations, a large non-linearity is needed to make the transitions individually addressable. 

\begin{figure}
    \centering
    \includegraphics[width = \linewidth]{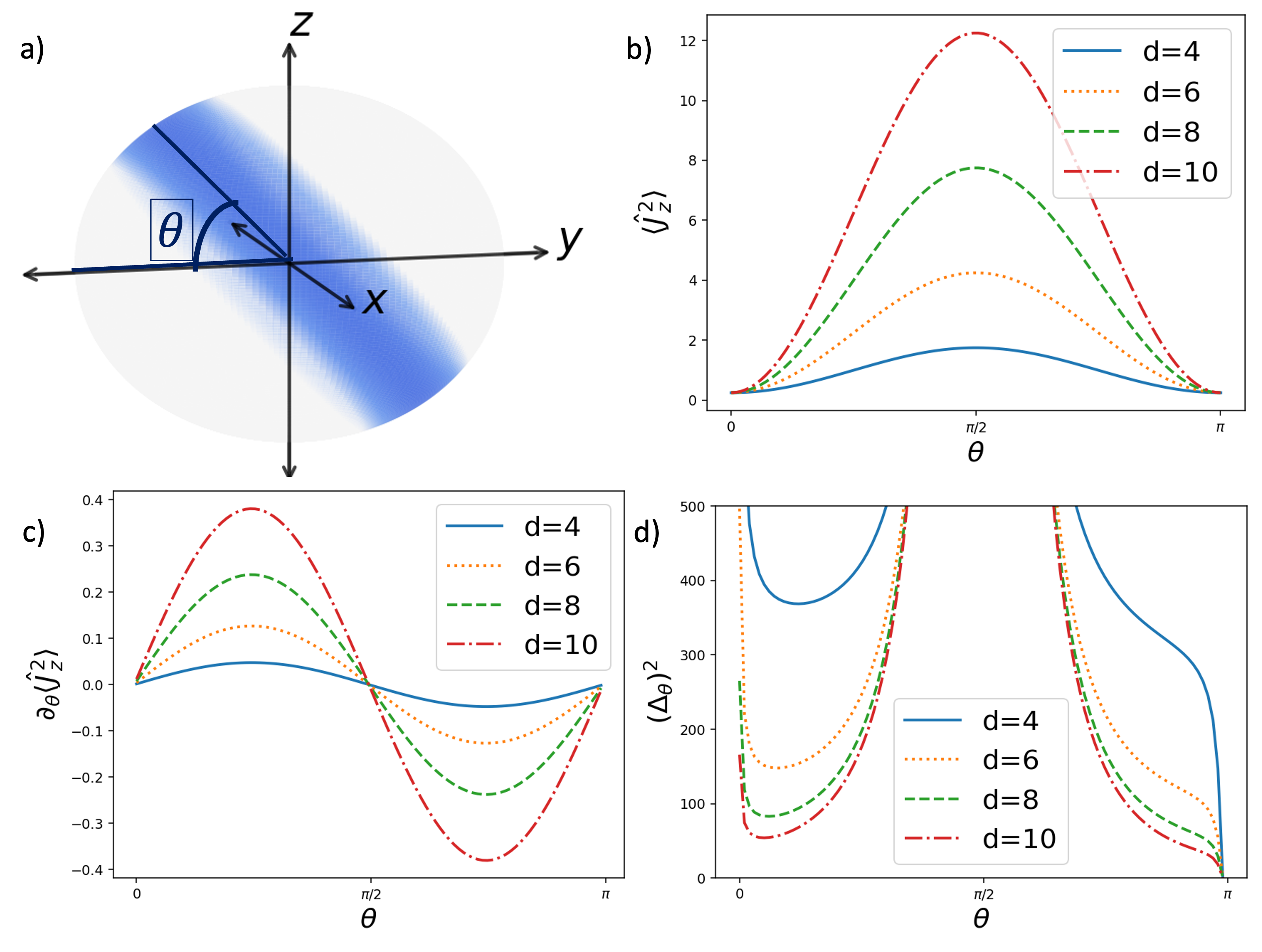}
    \caption{Parameter estimation using Dicke-like state on a qudit: a) A parameter $\theta$ is encoded by rotating a Dicke-like state about an axis perpendicular to the state.  b) Expectation value of $\hat J_z^2$. c) Rate of change of the expectation value with respect to the parameter $\theta$. d) Precision calculated using the error propagation formula, which diverges as $\partial_\theta\langle\hat J_z^2\rangle\to 0$. We note that $\Delta\theta$ decreases with the qudit dimension, and is proportional to $\nicefrac{1}{d^2}$, as can be seen by comparing $d=4$ and $d=8$. }
    \label{fig:dicke}
\end{figure}

\paragraph*{Phase encoding and readout:} Since a Dicke state is unchanged under $z$ rotations, we use a spin rotation about an axis perpendicular to the state, as shown in Fig.~\ref{fig:dicke}(a). The quantum Fisher information is given by 
\begin{equation}
    F_\text Q (\ket{i},\hat J_x) = 2i(d-1) - 2i +\frac{(d-1)}{2},
\end{equation}
which shows the standard quantum limit if $i \in \{0,d-1\}$ and shows a gain in precision for other values, being maximum when $i=\nicefrac{(d-1)}{2}$.
The rotation angle cannot be read out from measuring $\hat J_z$ due to the symmetric nature of the Dicke state. Instead, we use the expectation value of $\hat J_z^2$, which increases as the state is rotated about the $x$ axis, until a $\frac{\pi}{2}$ rotation and decreases thereafter.

\paragraph*{Parameter estimation:}
We calculate the expectation value of the observable $\hat J_z^2$ to estimate the parameter $\theta$, as shown in Fig.~\ref{fig:dicke}(b).
For calculating the precision, first, we calculate the variance and the rate of change of the expectation value with respect to the parameter $\theta$ (Fig.~\ref{fig:dicke}(c)), and then use Eq.~\ref{eq:precision}, with results shown in Fig.~\ref{fig:dicke}(d).
We note that precision diverges as $\partial_\theta\langle\hat J_z^2\rangle\to 0$ and  $\Delta \theta^2$ decreases quadratically with the qudit dimension, demonstrating Heisenberg limited scaling. 
The example of a Dicke-like state explictly uses an SU(2) symmetry for encoding a parameter on a qudit; next, we show an example which works without this assumption.

\subsection{GHZ-like state}\label{sec:GHZ}
\paragraph*{State preparation:}
The analogous of the GHZ state on a qudit is $\ket{0}+\ket{d-1}$ (unit norm implied), similar to the superposition of the highest and the lowest weight state in the Dicke basis,  given by $\ket{J,-J}+\ket{J,J}$, which can also be thought of as a spin cat state. 
In multi-particle systems, GHZ states can be prepared using two-qubit gates that entangle two atoms at a time and can be sequentially applied for multi-qubit entanglement. However, unlike multi-particle systems, a qudit cannot be partitioned into such subsystems; hence, two-qubit gates are not accessible. Instead, Givens rotations that selectively address a two-level subspace within a larger Hilbert space offer a solution for control of qudits.  To prepare a GHZ-like state on a qudit, we start with the qudit in the classical state $\ket{0}$ and apply a $\frac{\pi}{2}$-rotation directly coupling the $\ket{0}$ level to $\ket{d-1}$, resulting in $(\ket{0}+\ket{d-1})/\sqrt{2}$. Such a transition can be achieved either by multi-photon driving or using a sequence of pulses: one $\pi/2$ pulse between states $\ket{0}$ and $\ket{1}$, and then, $\pi$-pulses that shift the population from the $\ket{1}$ state to the $\ket{d-1}$ state, resulting in the GHZ-like state  $\ket{0}+\ket{d-1}$. In presence of non-linearities, this state can also be prepared using one-axis twisting.

\paragraph*{Phase encoding:}
We encode an unknown parameter $\theta$ into the relative phase of the GHZ-like state of a qudit. 
This could take form of interaction with a the magnetic or electric field of unknown strength $\beta$ for time $\tau$, that encodes $\theta=\beta\tau$.
The phase encoding operator $\hat P$ is represented by 
\begin{equation}
    \hat P = \text{diag}(0, 1, \ldots, d-1),
\end{equation}
such that applying $e^{i\theta \hat P}$ causes the transformation $\ket{0}+\ket{d-1}\to \ket{0}+e^{i\theta (d-1)}\ket{d-1}$.
We can note that the relative phase picked up by the GHZ-like state is enhanced by a factor of $(d-1)$, that would lead to metrological gain.

\paragraph*{Readout:}
The unknown parameter $\theta$ cannot be measured by directly measuring $\hat{P}$, which only reveals the average energy level, we describe an interferometric scheme for inducing phase-dependent population shifts to readout the relative phase.
In interferometry, the information about the relative phase is accessed by superimposing  two states, that combine in a constructive or destructive way depending on the phase difference, which can be inferred from population measurements of the final state.
To recombine the two components of the GHZ-like state of a qudit, we apply a $\pi$ rotation between the $\ket{d-1}$ and $\ket{1}$ states, such that $\ket{0}+e^{i\theta (d-1)}\ket{d-1}\to \ket{0}+e^{i\theta (d-1)}\ket{1}$. Such a transformation can be obtained either by multi-photon transition between the two levels or a sequence of $\pi$ pulses coupling the two levels via intermediate levels.

After recombining the two components of the GHZ-like state, the qudit is a coherent state, shifted by an angle $\theta(d-1)$ from the $x$ axis of the Bloch sphere of the two-level subspace $\ket{0}$ and $\ket{1}$. We apply a $\nicefrac{\pi}{2}$ rotation between $\ket{0}$ and $\ket{1}$ that transforms the state as 
\begin{equation}
\begin{split}
    \ket{0}+& e^{i\theta (d-1)}\ket{1} \to \\
    &\left(\ket{0}+i\ket{1}\right)+ie^{i\theta (d-1)}\left(\ket{0}-i\ket{1}\right)\\
    = &\left(1+e^{i\theta(d-1)}\right)\ket{0}+ i\left(1-e^{i\theta (d-1)}\right)\ket{1}.
\end{split}    
\end{equation}
Substituting $x=\theta(d-1) $, the coefficient of $\ket{0}$ is 
\begin{equation}
\begin{split}
    (1+e^{ix})&= (2\cos^2\frac{x}{2}+2i\sin\frac{x}{2}\cos{\frac{x}{2}}) \\
    & = 2\cos\frac{x}{2}\left(\cos{\frac{x}{2}}+i\sin{\frac{x}{2}}\right) = 2e^{i\frac{x}{2}}\cos{\frac{x}{2}}.
\end{split}
\end{equation}
Similarly, the coefficient of $\ket{1}$ is $i(1-e^{ix})= 2e^{i\frac{x}{2}}\sin{\frac{x}{2}} $. 
Thus, the final state after normalization is given by 
\begin{equation}\label{eq:finalcat}
    \psi = \left(\cos{\frac{\theta(d-1)}{2}}\ket{0}+\sin{\frac{\theta(d-1)}{2}}\ket{1}\right),
\end{equation}
ignoring the global phase. 
The population of the final state depends on the relative phase of the GHZ-like probe, which allows us to estimate the unknown parameter $\theta$.
Specifically, the population of $\ket{0}$ oscillates as $\cos^2\left((d-1)\theta/2\right)$, where the periodicity increases with the qudit dimension $d$, which leads to enhanced estimation of $\theta$.
Alternatively, in metrological protocols with GHZ states, the phase is inferred by measuring the qubit states in a complementary basis~\cite{PSO18RMP} or using parity oscillations, which could also be done for qudits.

\begin{figure}[h]
    \centering
    \includegraphics[width = \linewidth]{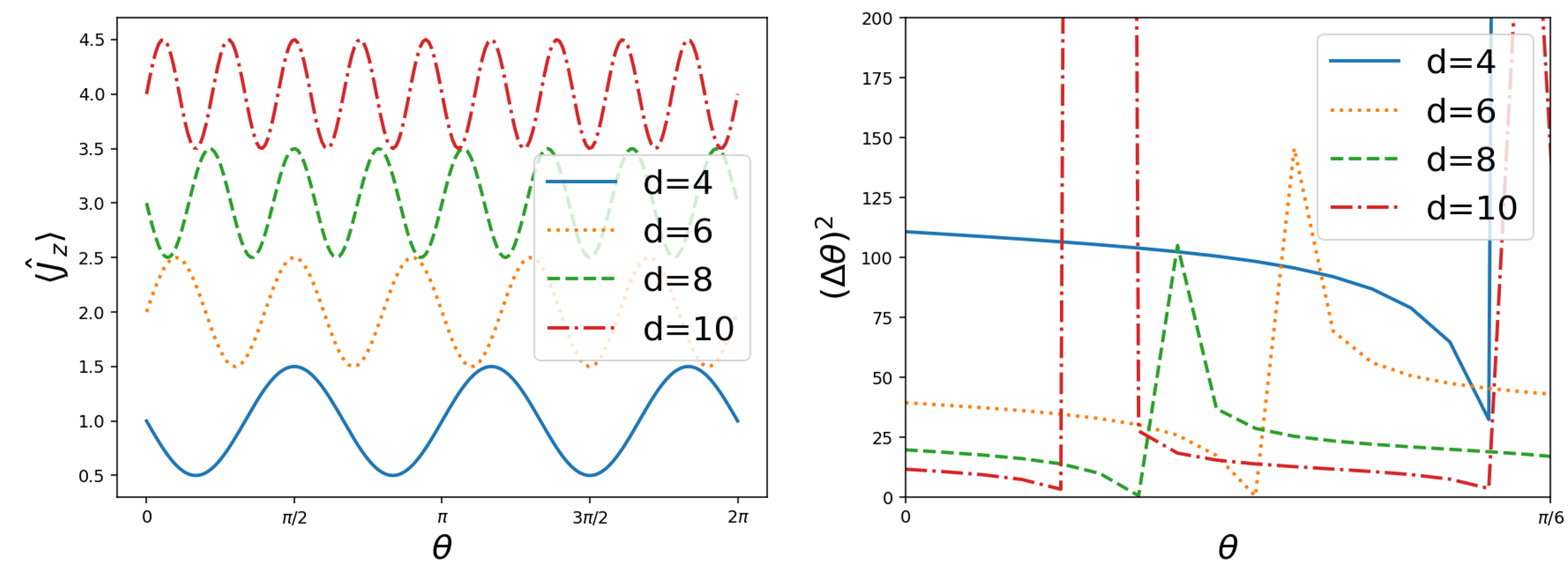}
    \caption{Parameter estimation using a GHZ-like state on a qudit, where a parameter $\theta$ is encoded by in the relative phase of the two components. a) Expectation value of $\hat J_z$ following a generalized interferometry protocol. d) Precision calculated using the error propagation formula, which diverges as $\partial_\theta\langle\hat J_z^2\rangle\to 0$. We note that $\Delta\theta$ decreases with the qudit dimension, and is proportional to $\nicefrac{1}{d^2}$, as can be seen by comparing $d=4$ and $d=8$.  }
    \label{fig:catestimate}
\end{figure}

\paragraph*{Parameter estimation:}
We calculate the expectation value of the observable $\hat P$ to estimate the parameter $\theta$, which acts on eigenstates as $\hat{P}\ket{i}= i\ket{i}$ for $i\in[0,d-1]$.
The mean value $\langle{\hat P}\rangle= \sin^2{\frac{\theta(d-1)}{2}}$ for the state in Eq.~\ref{eq:finalcat}.
Figure~\ref{fig:catestimate}(a) shows the oscillation of $\langle\hat J_z\rangle =\langle\hat P\rangle+\nicefrac{(d-1)}{2} $, where the increasing frequency of oscillation indicates the enhanced phase sensitivity as the qudit dimension increases.
For calculating the precision, first, we calculate the variance using  $\langle{\hat P^2}\rangle = \sin^2{\frac{(d-1)\theta}{2}}$
and $(\Delta \hat P)^2 := \langle{\hat P^2}\rangle - (\langle{\hat P}\rangle)^2$,
such that  
\begin{equation}\label{eq:var}
    (\Delta \hat P)^2 = \frac{\sin^2((d-1)\theta)}{4},
\end{equation}
using $\sin^2(a)-\sin^4(a) = \sin^2(2a)/4$. The above expression forms the numerator for the estimation precision Eq.~\ref{eq:precision}.
For the denominator, we calculate the rate of change of $\langle{\hat P}\rangle$ with respect to the phase $\theta$, such that
\begin{equation}\label{eq:dIz}
    |\partial_\theta\langle \hat P\rangle|^2 = \left(\frac{(d-1)} {2}\sin (d-1)\theta\right)^2.
\end{equation}
Substituting Eqs.~\ref{eq:var} and \ref{eq:dIz} into Eq.~\ref{eq:precision}, we find the precision to be
\begin{equation}
    (\Delta\theta)^2  = \frac{\sin^2((d-1)\theta)}{(d-1)^2\sin^2((d-1)\theta)} \xrightarrow{\theta\to0} \frac{1}{(d-1)^2},
\end{equation}
where we can note that the precision increases linearly with $d-1$.
Figure~\ref{fig:catestimate}(b)  shows the precision $\Delta\theta$ vs $\theta$ obtained from numerical simulations, which diverges as $2I\theta\to n\pi$ for $n\in\mathbb{Z}$.

\section{Effect of Decoherence}\label{sec:decoherence}

We study the scaling of the coherence of a GHZ-like state with respect to the Hilbert space dimension of a qudit and consider dephasing of the state $\ket{0}+\ket{d-1}$  for different values of $d$.
The density matrix of such a state consists only of four elements: two diagonal elements corresponding to the population in $\ket{0}$ or $\ket{d-1}$ and two off-diagonal components that quantify the coherence of the superposition state.
We write density matrix of a GHZ-like state under decoherence to be
\begin{equation}
\begin{split}
    \rho^{\text {GHZ}} = \frac{1}{2}\Big[ &\ket{0}\bra{0}+\ket{d-1}\bra{d-1}\\
    &+ \ket{0}\bra{d-1}e^{-(d-1)^2\gamma t}\cos(d-1)\theta \\
    &+ \ket{d-1}\bra{0}e^{-(d-1)^2\gamma t}\sin(d-1)\theta \Big],
\end{split}
\end{equation}
where $\gamma$ is the rate of dephasing and $t$ the interaction time of the probe.
Thus, QFI can be calculated used Eq.~\ref{eq:QFI-mixed} as
\begin{equation}
    F_\text Q^{\text {GHZ}} = (d-1)^2e^{-(d-1)^2\gamma t}.
\end{equation}
We can note that QFI decreases with the dimension $d$ and results from a linear decrease in coherence time as the dimension of a qudit increases.
Such a trade-off between lifetime and the macroscopicity of a GHZ-like state is similarly found in other systems such as harmonic oscillators~\cite{KTVJ19PRL} and maximally entangled GHZ states~\cite{HMP+97PRL}.
Metrological power is given by
\begin{equation}
    \mathcal{W}(\rho^{\text {GHZ}} ) = \max \left[(d-1)e^{-(d-1)^2\gamma t}-1, 0\right],
\end{equation}
which is positive for small values of $d$.
Since metrological power is bounded by non-classicality (Eq.~\ref{eq:relationWN}), a non-zero $\mathcal{W}$ also serves as a witness of non-classicality.

\section{Conclusion}~\label{sec:conclusion}
We introduced metrological schemes for parameter estimation that achieve the Heisenberg limit using a single $d$-dimensional qudit, with precision scaling as $d-1$, and show in a linear advantage in using qudits.
The key advantage in our scheme arises from the use of non-classical states on a qudit, which  have a large quantum Fisher information, and can serve as a resource for metrology, shown explicitly through measures of non-classicality and metrological gain introduced here.
Our work opens the possibility for realizing quantum-enhanced parameter estimation with independent probes, without the use of entanglement, that could be utilized on several platforms such as nuclear spins~\cite{AMJ+20Nature}, atomic spins~\cite{CBE+18NatComm}, NV centres~\cite{DFD+11NatPhys} and superconducting circuits~\cite{SZS+18PRA}, that host a high-dimensional Hilbert space.
Possible extensions of this work include multi-parameter estimation~\cite{LHXX20JoPA} using qudits, and protecting the encoded parameter from decoherence through information scrambling~\cite{LCS+23Science}.

\section*{Acknowledgements}
I thank Abhijeet Alase, Barry C. Sanders and Andrea Morello for useful discussions, and NSERC for funding this work. 
\bibliography{main}
\end{document}